\documentclass[12pt]{article}

\usepackage{latexsym}
\usepackage{amssymb}
\usepackage{amsbsy}
\usepackage{epsfig}
\usepackage{graphics,color}
\usepackage{a4}

\newcommand{\be}{\begin{eqnarray}}
\newcommand{\ee}{\end{eqnarray}}
\newcommand{\bi}{\begin{itemize}}
\newcommand{\ei}{\end{itemize}}

\newcommand{\vck}{{\bf k}}
\newcommand{\vcx}{{\bf x}}
\newcommand{\mpl}{M_{\rm pl}}

\begin{document}

\title{
\vskip -100pt
{\begin{normalsize}
\vskip  100pt
\end{normalsize}}
{\bf\Large Quantum field thermalization in expanding backgrounds}
\author{
\addtocounter{footnote}{2}
Anders Tranberg\thanks{email: anders.tranberg@oulu.fi}\\
 {\em\normalsize Department of Physical Sciences, University of Oulu} \\
   {\em\normalsize P.O. Box 3000, FI-90014 Oulu, Finland}
}
}
\date{\today}
\maketitle

\begin{abstract}
The 2PI effective action formalism for quantum fields out of equilibrium is set up in an expanding (Friedmann-Robertson-Walker) background. We write down and solve the evolution equations for a $\varphi^4$ model at $\mathcal{O}(\lambda^2)$ in a coupling expansion. We comment on issues of renormalization, lattice discretization and the range of applicability of the approach. A number of example calculations are presented, including thermalization and (p)reheating. Generalizations to more complicated systems and applications are discussed.
\end{abstract}

% SECTION: 2PI IN EXPANDING BACKGROUND

\section{Introduction\label{sec:introduction}}

In recent years, significant attention has been drawn to the process of thermalization of quantum fields. Quantitative description of the physics of the very early Universe and of heavy-ion collision experiments requires an understanding of the real-time dynamics of quantum fields at finite energy density, but out of equilibrium. 

 One very promising development is the application of the 2PI-formalism \cite{Cornwall:1974vz,Berges:2000ur,Berges:2001fi}, which allows the derivation and explicit numerical solution of a set of equations of motion for the mean field and propagator in the full quantum theory. This is realized through the truncation of a controlled diagram expansion in terms of 2PI diagrams. Already at next-to-leading order (NLO) in either a coupling or $1/N$ expansion, interacting systems exhibit equilibration, effective dissipation and thermalisation to the {\it quantum} equilibrium state \cite{Berges:2000ur,Aarts:2002dj,Berges:2002wr,Juchem:2003bi,Berges:2004ce,Arrizabalaga:2005tf,Lindner:2005kv}. 

The physics of the early Universe is described not in a static background, but in expanding space. This is often approximated by a homogeneous, flat Friedmann-Robertson-Walker (FRW) space-time, parametrized by the scale factor $a(t)$ of the metric $ds^2=dt^2-a^2(t)d{\bf x}^2$. In some cases, expansion can be neglected when the time scale of a phenomenon is very short on the time scale of expansion. But in general, and of course in principle, expansion should be included in the description of early Universe physics. 

Many processes at high temperature or energy density are well described by the classical approximation, where a Monte-Carlo sample of initial field configurations are evolved using Hamiltonian equations of motion (see \cite{Smit:2005vp} for a brief review). The observables of interest are then averages over this classical ensemble. The approximation must however break down eventually, as classical fields equilibrate to a classical equilibrium, which suffers from the Rayleigh-Jeans problem: in the continuum temperature will go to zero, and on the lattice it will be cut-off dependent. 

In this paper, we set out the 2PI formalism in a FRW space-time, and solve the resulting equations numerically for some example applications. A number of studies have been carried out in this context in the Hartree approximation \cite{Boyanovsky:1993xf,Khlebnikov:1996mc,Khlebnikov:1996wr,Boyanovsky:1996fz,Boyanovsky:1997cr} (which is also leading order (LO) in a 2PI coupling expansion), and even before that, the formalism was set out in \cite{Calzetta:1986ey,Calzetta:1986cq}. Recently, attempts have been made to partly include the effect of an expanding background for specific applications also at NLO \cite{Rajantie:2006gy,Aarts:2007qu,Aarts:2007ye}. 

Two main issues present themselves. Firstly, as we are discretizing the system on a finite co-moving lattice, there is only a finite number of momentum modes available, and as the lattice expands in time these will be redshifted towards the IR in physical units. This means that the physical cut-off changes in time. Therefore, there is a limit on how many e-folds one can run the simulation before running out of ``dynamical range''. In practice, this means that at some point discretization errors become important, and results can no longer be relied upon. As a result, reliably simulating cosmological inflation proper is a daunting task, as the Universe expands many e-folds. Still, most of the inflationary stage is often well described by semi-analytical tools and the slow-roll approximation, and only the couple of e-folds around the end of inflation, reheating and the transition to radiation domination requires numerical treatment. Post-inflationary phenomena typically only span a few e-folds.

Secondly, since we are doing quantum physics, the theory has to be renormalized. In particular, the energy density which enters in the semi-classical Friedmann equation (see below) needs appropriate counterterms. Fortunately, features of the 2PI formalism include that it is renormalizable at any level of diagram truncation  \cite{vanHees:2001ik,Blaizot:2003br,Berges:2004hn,Berges:2005hc,Arrizabalaga:2006hj}, and that there is a similarly truncated energy density which is conserved. Hence by introducing (scale-factor dependent) counterterms for the energy density, mass and couplings, we can in principle cancel all divergences, and construct a well-behaved Friedmann equation.

We study a self-interacting real scalar, and go to NLO ($\mathcal{O}(\lambda^2)$) in a 2PI coupling expansion. By showing how to apply the procedure in practice, we expect it will be clear how to generalize to more complicated systems. In the conclusions we point out some issues, applications and ways of refining the approach.

% SUBSECTION: SETUP

\subsection{Setup\label{sec:setup}}

We are concerned with a single scalar field with $\varphi^4$ interaction. The action is
\be
S= \int dt\,d^{3}\vcx\, a^{3}(t)\bigg[ \frac{1}{2}(\partial_t\varphi)^{2}-\frac{1}{2a^2(t)}(\partial_{\bf x}\varphi)^{2}-\frac{1}{2}m^2\varphi^2-\frac{\lambda}{24}\varphi^4\bigg],
\label{eq:action1}
\ee
written in terms of co-moving spatial coordinates {\bf x}. $a(t)$ is the scale factor, and we assume $a(0)=1$. Correspondingly, we will consider co-moving and physical momenta, denoted $\vck$ and $\tilde\vck=\vck/a(t)$, respectively.

The evolution of the scale factor is in turn given by the Friedmann equation in terms of the Hubble rate $H$\footnote{We use $\dot{a}$ to denote $\partial_t a(t)$. $a'$ will denote $\partial_\eta a(\eta)$, with $\eta$ conformal time.},
\be
H^2(t)=\frac{1}{3M_{\rm pl}^2}\langle T^{00}(t)\rangle_{\rm ren}, \qquad H(t)=\frac{\dot{a}(t)}{a(t)}.
\label{eq:friedmann}
\ee
Here we equate a classical quantity on the left-hand side to a quantum expectation value on the right-hand side. This only makes sense when the energy density is appropriately renormalized, an issue we will return to below.

The system can be recast in comoving (conformal) time $\eta$, with $dt= a(\eta)d\eta$ and we can rescale the field\footnote{The equations of motion will be solved in terms of the ``conformal'' field $\phi(\eta)$, but results converted back to the ``physical'' field $\varphi(t)$.} $\varphi(x)=\phi(x)/a(\eta)$, in which case the action becomes
\be
S= \int d\eta\,d^{3}\vcx\,\bigg[ \frac{1}{2}(\partial_\eta \phi-\mathcal{H}\phi)^{2}-\frac{1}{2}(\partial_{\bf x} \phi)^2-\frac{1}{2}a^2(\eta)m^2\phi^2-\frac{\lambda}{24}\phi^4\bigg].
\label{eq:action2}
\ee
We have introduced a new ``comoving Hubble rate'', $\mathcal{H}=a'/a=aH$. In terms of the canonical momentum $\pi=\partial_\eta \phi-\mathcal{H}\phi$, the corresponding Friedmann equation is \footnote{The right-hand side is not the Hamiltonian corresponding to the action (\ref{eq:action2}), but the Hamiltonian of (\ref{eq:action1}), written in terms of the rescaled fields.}
\be
\frac{(a')^2}{a^4}=\frac{1}{3a^4M_{\rm pl}^2}\langle\bigg[ \frac{1}{2}\pi^{2}+\frac{1}{2}(\partial_{\bf x} \phi)^2+\frac{1}{2}a^2(\eta)m^2\phi^2+\frac{\lambda}{24}\phi^4\bigg]\rangle.
\label{eq:friedmann2}
\ee

In passing, it is useful to recall the classical equation of motion
\be
\left[\partial_\eta^2-\partial_\vcx^2-\frac{a''}{a}+a^2m^2+\frac{\lambda}{6}\phi^2(x)\right]\phi(x)=0.
\label{eq:classical}
\ee
The classical approximation amounts to generating a set of random initial conditions, solve for the evolution using (\ref{eq:classical}) together with the Friedmann equation and then to average observables over initial conditions. In addition to using approximate dynamics, also the classical averaging procedure is different from (\ref{eq:friedmann2}) in that the Hubble rate, and hence $a(t)$, is derived from each individual initial condition rather than the average energy density. This could be resolved by simulating all initial conditions simultaneously using a common $a(t)$ determined through the ensemble averaged energy density. Still, it would be a classical average rather than a quantum one.

% SECTION: 2PI EQUATIONS

\section{The 2PI formalism\label{sec:2PI}}

% SUBSECTION : EQUATIONS OF MOTION at Order lambda squared

\subsection{Equations of motion at $\mathcal{O}(\lambda^2)$\label{sec:2PIeq}}

We will not review the 2PI formalism in detail here, but refer to literature on the subject (for instance \cite{Berges:2004yj} and references therein). 
Applying the 2PI formalism to the action (\ref{eq:action1}), while treating the scale factor $a(t)$ as an external field, we find the equations of motion for the homogeneous mean field $\bar\phi(\eta)=\langle\phi(x)\rangle$ and the propagator\footnote{The fields are defined along the Keldysh contour $\mathcal{C}$.} 
\be
\langle T\phi(x)\phi(y)\rangle-\bar\phi(\eta)\bar\phi(\eta')=F(\eta,\eta',{\bf x-y})-\frac{i}{2}\rho(\eta, \eta',{\bf x-y}){\rm \,sign}_{\mathcal{C}}(\eta-\eta'),
\ee 
where $F$ and $\rho$ are real. 

The general form is
\be
\left[\partial_\eta^2+M^2_\phi(\eta)\right]\bar\phi(\eta)&=&-\int_0^\eta d\eta' \int d^3{\bf x}\,\Sigma_\phi(\eta,\eta',{\bf x})\bar\phi(\eta'),\\
\left[\partial_\eta^2+k^2+M^2(\eta)\right]F(\eta,\eta',{\bf k})&=&-\int_{0}^{\eta}d\eta'' \Sigma_\rho(\eta,\eta'',{\bf k})F(\eta'',\eta',{\bf k})\nonumber\\
&&+\int_{0}^{\eta'}d\eta'' \Sigma_F(\eta,\eta'',{\bf k})\rho(\eta'',\eta',{\bf k}),\\
\left[\partial_\eta^2+k^2+M^2(\eta)\right]\rho(\eta,\eta',{\bf k})&=&-\int_{\eta'}^{\eta}d\eta'' \Sigma_\rho(\eta,\eta'',{\bf k})\rho(\eta'',\eta',{\bf k}).
\ee
These are known as the Kadanoff-Baym \cite{Baym:1962sx} or real-time Schwinger-Dyson equations. They include a set of self-energies $M^2_\phi$, $M^2$, $\Sigma_\phi$, $\Sigma_F$, $\Sigma_\rho$, to be calculated and integrated up with the correlators for all past time. This makes 2PI simulations numerically challenging, although one should remember that 2PI simulations do not require statistical averaging, since the variables $F$, $\rho$, $\bar{\phi}$ are already the full correlators. 

Throughout, we use Fourier transforms in comoving coordinates and assume homogeneity, so that\footnote{We use a unifying notation for momentum integrals in the continuum $\int_\vck=\int\frac{d^3\vck}{(2\pi)^3}$ and on the lattice $\int_\vck=\frac{1}{V}\sum_\vck$.} 
\be
\langle\phi(\eta,{\bf x})\phi(\eta',{\bf y})\rangle=\int_\vck\langle\phi_{\vck}(\eta)\phi_{-\vck}(\eta')\rangle e^{i{\bf k}({\bf x-y})}
.
\ee

In the 2PI formalism, the self-energies are determined from truncations of a (2PI-)diagram expansion. At NLO in a coupling expansion \cite{Berges:2000ur,Arrizabalaga:2005tf}, 
\be
\label{eq:Meff}
M^2_\phi(\eta)&=&-\frac{a''(\eta)}{a(\eta)}+a^2(\eta)m^2+\frac{\lambda}{6}\left(\bar\phi^2(\eta)+3F(\eta,\eta,{\bf x=0})\right),\\
M^2(\eta)&=&-\frac{a''(\eta)}{a(\eta)}+a^2(\eta)m^2+\frac{\lambda}{2}\left(\bar\phi^2(\eta)+F(\eta,\eta,{\bf x=0})\right),
\ee
\be
\Sigma_F(\eta,\eta',{\bf x})&=&-\frac{\lambda^2}{2}\bar\phi(\eta)\left(F^2(\eta,\eta',{\bf x})-\frac{1}{4}\rho^2(\eta,\eta',{\bf x})\right)\bar\phi(\eta')\nonumber\\
&&-\frac{\lambda^2}{6}F(\eta,\eta',{\bf x})\left(F^2(\eta,\eta',{\bf x})-\frac{3}{4}\rho^2(\eta,\eta',{\bf x})\right),\\
\Sigma_\rho(\eta,\eta',{\bf x})&=&-\frac{\lambda^2}{2}\bar\phi(\eta)2F(\eta,\eta',{\bf x})\rho(\eta,\eta',{\bf x})\bar\phi(\eta')\nonumber\\
&&-\frac{\lambda^2}{6}\rho(\eta,\eta',{\bf x})\left(3F^2(\eta,\eta',{\bf x})-\frac{1}{4}\rho^2(\eta,\eta',{\bf x})\right),\\
\Sigma_\phi(\eta,\eta',{\bf x})&=&-\frac{\lambda^2}{6}\rho(\eta,\eta',{\bf x})\left(3F^2(\eta,\eta',{\bf x})-\frac{1}{4}\rho^2(\eta,\eta',{\bf x})\right).
\ee
The only difference to the Minkowski-space equations is in the $a(\eta)$-dependence of the effective masses. This is the standard free-field dependence, also appearing in the classical equation of motion (\ref{eq:classical}). Note however that the time coordinate is the conformal one and that we still consider the rescaled field $\phi$. Physical time is therefore
\be
t-t_0=\int^\eta_{\eta_0} a(\eta) d\eta,
\label{eq:phystime}
\ee
and the physical correlators are
\be
\langle\varphi(x)\rangle&=&\bar{\varphi}(x)=\frac{1}{a(\eta)}\bar{\phi}(x),\\
\langle T\varphi(x)\varphi(y)\rangle&=&\frac{1}{a^2(\eta)}\langle T\phi(x)\phi(y)\rangle,\\
\langle T\partial_t\varphi(x)\partial_t\varphi(y)\rangle&=& \frac{1}{a^4(\eta)}\langle T[\left(\partial_\eta-\mathcal{H}\right)\phi(x)][\left(\partial_\eta-\mathcal{H}\right)\phi(y)]\rangle.
\ee
For $m=0$ and with non-minimally ``conformal'' coupling to curvature (an additional term $\frac{1}{6}\varphi^2 R$ in the action), no trace remains of the expansion in the equation of motion, and physical processes proceed exactly as in Minkowski space, except that the time coordinate is ``stretched'' through (\ref{eq:phystime}). We use minimal coupling to gravity and non-zero mass, leading to deviations from conformal behaviour.

% SUBSECTION: RENORMALIZATION

\section{Renormalization\label{sec:2PIrenorm}}

Renormalization in an expanding space-time requires additional counterterms compared to the Minkowski case, as operators involving the metric (in this case, the scale factor $a(t)$) emerge in the effective action. We will adopt the approach of \cite{Anderson:2005hi}, where the energy density is renormalized by subtracting a contribution corresponding to the adiabatic vacuum solution in the background defined by $a(t)$. This vacuum solution can be solved for order by order\footnote{For brevity, we write $\dot{a}^n$ to mean all combinations of $n$ time derivatives. For $n=2$ we have terms and $\dot{a}^2$ and $\ddot{a}$, etc. Similarly $\mathcal{O}(H^2)$ includes $\mathcal{O}(\dot{H})$.} in derivatives of $a(t)$ using a WKB-type ansatz: at leading order ($\dot{a}^0$) all quartic divergences are canceled, at the next order ($\dot{a}^2$) also quadratic divergences and finally at order $\dot{a}^4$ the logarithmic divergences. 

In Minkowski space, a full-fledged 2PI renormalization procedure has been developed \cite{vanHees:2001ik,Blaizot:2003br,Berges:2004hn,Berges:2005hc,Arrizabalaga:2006hj}, which allows a proper continuum limit to be taken. Since we are not intending to go to the continuum, we here take a somewhat simpler approach, and renormalize the equations of motion and the energy density in the LO (Hartree) approximation by a mass and energy counterterm only \cite{Arrizabalaga:2005tf}. The mass counterterm is calculated using the same WKB vacuum solution as for the energy density. From a numerical viewpoint, logarithmic divergences are very small indeed, and so we restrict ourselves to renormalizing the energy density to this level of precision, i.e. expanding the WKB solution only to second order in $\dot{a}$.

\subsection{Counterterms\label{sec:counterterms}}

The energy density is, in the 2PI-LO approximation ($\tilde{F}$ is the correlator for the original $\varphi$ field)
\be
\langle T^{00}(t)\rangle_{\rm ren}&=&
\bigg[ \frac{1}{2}(\partial_t \bar{\varphi})^{2}
+\frac{1}{2}\partial_{t}\partial_{t'}\tilde{F}(t,t',{\bf 0})_{t=t'}
\nonumber\\
&&\frac{1}{2a^2(t)}\partial_{\bf x}\partial_{\bf x'}\tilde{F}(t,t,{\bf x-x'})_{\vcx=\vcx'}
+\frac{1}{2}m^2_b\left(\bar{\varphi}^2(t)+\tilde{F}(t,t,{\bf x=0})\right)+\nonumber\\
&&\frac{\lambda}{24}\left(\bar{\varphi}^4(t)+6\bar{\varphi}^2(t)\tilde{F}(t,t,{\bf x=0})+3\tilde{F}^2(t,t,{\bf x=0})\right)-\delta T^{00}
\bigg].
\ee
We have introduced a bare mass $m_b(t)$ and an energy counterterm $\delta T^{00}(t)$ to be determined. These are time dependent, but only through $a(t)$. The energy can be written in a more suggestive form as
\be
&&\langle T^{00}(t)\rangle_{\rm ren}=\Bigg[ \frac{1}{2}(\partial_t \bar{\varphi})^{2}+\frac{1}{2}\left(m_b^2+\frac{\lambda}{2}\tilde{F}(t,t,{\bf x=0})\right)\bar{\varphi}^2(t)+\frac{\lambda}{24}\bar{\varphi}^4(t)+\nonumber\\
&&\int_\vck\frac{1}{2}\left(\partial_{t}\partial_{t'}\tilde{F}(t,t',{\bf k})_{t=t'}+\left(\frac{\vck^2}{a^2(t)}+m_b^2+\frac{\lambda}{2}\tilde{F}(t,t,{\bf x=0})+\frac{\lambda}{2}\bar{\varphi}^2(0)\right)\tilde{F}(t,t,{\bf k})\right)\nonumber\\
&&-\frac{\lambda}{4}\bar{\varphi}^2_0\tilde{F}(t,t,{\bf x=0})-\frac{\lambda}{8}\tilde{F}^2(t,t,{\bf x=0})-\delta T^{00}(t)
\bigg].
\ee
where $\bar{\varphi}_0=\langle\varphi(t=0)\rangle$. The LO equations of motion read
\be
\left[\partial_t^2+3H(t)\partial_t+\tilde\omega^2_\vck(t)-\frac{\lambda}{3}\bar{\varphi}^2(t)\right]\bar{\varphi}(t)=0,\\
\left[\partial_t^2+3H(t)\partial_t+\tilde\omega^2_\vck(t)\right]\tilde{F}(t,t,\vck)=0,
\label{eq:LOeom2}
\ee
with
\be
\label{eq:LOeom1}
\tilde\omega_\vck^2(t)=\vck^2/a^2(t)+m^2_b+\frac{\lambda}{2}\tilde{F}(t,t,{\bf x=0})+\frac{\lambda}{2}\bar{\varphi}^2(t).
\ee

We will renormalize in a vacuum to be determined below. Assuming we have such a vacuum, let us define
\be
\label{eq:massren}
m_b^2=m^2-\frac{\lambda}{2}\delta\tilde{F}_{\rm vac}(t,t,{\bf x=0}),
\ee
where $\delta\tilde{F}_{\rm vac}$ is a WKB approximation to the exact, time-dependent, vacuum solution $\tilde{F}_{\rm vac}(t,t,{\bf x=0})$.

Similarly, let us define
\be
\delta T^{00}&=&\delta T^{00}_{\rm free}+\delta T^{00}_{\rm int},\\
\delta T^{00}_{\rm free}&=&\int_k\frac{1}{2}\partial_{t}\partial_{t'}\delta\tilde{F}_{\rm vac}(t,t',{\bf k})_{t=t'}+\frac{1}{2}\left(\frac{k^2}{a^2(t)}+m^2+\frac{\lambda}{2}\bar{\varphi}^2(0)\right)\delta\tilde{F}_{\rm vac}(t,t,{\bf k})\\
\delta T^{00}_{\rm int}&=&
-\frac{\lambda}{4}\bar{\varphi}^2_0\delta\tilde{F}_{\rm vac}(t,t,{\bf x=0})-\frac{\lambda}{8}\delta\tilde{F}_{\rm vac}^2(t,t,{\bf x=0}).
\label{eq:enren}
\ee

We will be interested in evolving various, non-vacuum, initial conditions in real time, giving rise to some evolution of $a(t)$. In principle, we could then also solve the vacuum equation numerically, simultaneously in the background of that same $a(t)$, and self-consistently use the resulting $\tilde{F}_{\rm vac}$ for the counterterms of the non-vacuum simulation. 

Because the vacuum solution depends only on $a(t)$ and its time derivatives, and because the renormalised LO equations of motion in vacuum are similar to the free ones (insert (\ref{eq:massren}) into (\ref{eq:LOeom1} into\ref{eq:LOeom2})), we will instead solve a WKB equation for the field modes. In this way, the counterterms will simply be functions of $a(t)$, to be calculated at each time $t$. By including subsequent orders in WKB, we will be able to get a better and better approximation to $\tilde{F}_{\rm vac}$.

% SUBSECTION: ADIABATIC REGULARISATION

\subsection{Choosing the vacuum\label{sec:adiabaticregularisation}}

Following \cite{Anderson:2005hi}, we have for the original, un-rescaled field $\varphi$ a WKB ansatz
\be
\varphi_\vck(t)=a_\vck f_\vck(t)+a_\vck^\dagger f_\vck^{*}(t),\quad f_\vck(t)=\frac{1}{\sqrt{2\,a^3(t)\Omega_\vck(t)}}e^{-i\int^t\Omega_\vck(t')dt'}.
\ee
which is meant to satisfy the equation of motion for the free field modes $\varphi_\vck(t)$
\be
\left[\partial_t^2+3H(t)\partial_t+\tilde\omega^2_\vck(t)\right]\varphi_\vck(t)=0,
\ee
with 
\be
\tilde\omega_\vck^2(t)=k^2/a^2(t)+M^2,\qquad M^2=m^2+\frac{\lambda}{2}\bar{\varphi}^2(0).
\ee 
As mentioned, we include the possibility of having a non-zero initial mean field $\bar{\varphi}(0)$, in which case we include the initial value in the mass. 
As is argued in the Appendix, divergences associated with a time-dependent mass (and in this case, a varying mean field) are logarithmic, and so beyond the level of approximation aimed at here. 

$\Omega_\vck$ has to satisfy
\be
\label{eq:omegaeq}
\Omega^2_\vck=\tilde\omega^2_\vck-\frac{3}{2}\dot{H}-\frac{9}{4}H^2-\frac{1}{2}\frac{\ddot\Omega_\vck}{\Omega_\vck}+\frac{3}{4}\left(\frac{\dot\Omega_\vck}{\Omega_\vck}\right)^2.
\ee
At leading order in WKB, $\Omega_\vck=\tilde\omega_\vck$. By plugging this back into (\ref{eq:omegaeq}) we get the next order,
\be
\label{eq:omegadef}
(a\bar{\Omega}_\vck)^2=(a\tilde{\omega}_\vck)^2\left(1-\frac{a''/a}{(a\tilde{\omega}_\vck)^2}\left(1+\frac{1}{2}\left(\frac{aM}{a\tilde{\omega}_\vck}\right)^2\right)-\frac{\mathcal{H}^2}{(a\tilde{\omega}_\vck)^2}\left(\frac{1}{2}\left(\frac{aM}{a\tilde{\omega}_\vck}\right)^2-\frac{5}{4}\left(\frac{aM}{a\tilde{\omega}_\vck}\right)^4\right)\right).
\ee
where we have defined $\mathcal{H}=aH=a'/a$. $\bar{\Omega}_\vck$  defines our approximation to the (infinite order in WKB) vacuum. We define
\be
\delta\tilde{F}_{\rm vac}(t,t',{\bf x=0})=\int_\vck \langle\varphi_\vck^\dagger(t)\varphi_\vck(t')\rangle.
\ee
Using
\be
\dot{f}_\vck(t)=\left(-i\Omega_\vck-\frac{3}{2}H-\frac{1}{2}\frac{\dot{\Omega}_\vck}{\Omega_\vck}\right)f_\vck(t),
\ee
we have to this order and second order in $H$
\be
\label{eq:freevacE}
\delta T^{00}_{\rm free}
&=&\frac{1}{2a^4}\int_\vck a\tilde{\omega}_\vck\left(1+\frac{\mathcal{H}^2}{2(a\tilde{\omega}_\vck)^2}\left(1+\left(\frac{m}{\tilde{\omega}_\vck}\right)^2+\frac{1}{4}\left(\frac{m}{\tilde{\omega}_\vck}\right)^4\right)\right).
\ee
We also have
\be
\delta\tilde{F}_{\rm vac}(t,t,{\bf x=0})=\int_\vck \frac{1}{a^2}\frac{1}{2a\bar{\Omega}_\vck},
\ee
determining the bare mass (\ref{eq:massren}) and $\delta T^{00}_{\rm int}$ (\ref{eq:enren}).

With $m_b^2$, $\delta T^{00}_{\rm free}$ and $\delta T^{00}_{\rm int}$ as defined above, the renormalized energy density
has divergences $\propto C(t)\ln\Lambda$ in Minkowski space, where $C(t)$ is a function of mass dimension four of the effective time-dependent mass and its time derivatives. In FRW space, $C(t)$ can also depend on (time-derivatives of) $H$. In particular, in this approach the finite parts of the counterterms are chosen to cancel the 1-loop corrections to  $\mathcal{O}(H^2)$, and so effectively amount to the renormalization conditions, in terms of some momentum cut-off $\Lambda$,
\be
\langle T^{00}\rangle_{\rm ren}= 0+\mathcal{O}(C(0)\ln\Lambda),\qquad (\textrm{initially in vacuum, LO}),\\
m^2_{\rm ren}=m^2+\mathcal{O}(C(0)\ln\Lambda),\qquad (\textrm{initially in vacuum, LO}).
\ee

Using the  WKB solution with $\bar{\Omega}_\vck$ as an initial state, $\langle T^{00}\rangle_{\rm ren}$ is identically zero for zero mean field, or initially equal to the "tree-level" energy density for a non-zero mean field
\be
\langle T^{00}(0)\rangle_{\rm ren}&=&
\bigg[ \frac{1}{2}(\partial_t \bar{\varphi})^{2}(0)+\frac{1}{2}m^2\bar{\varphi}^2(0)+\frac{\lambda}{24}\bar{\varphi}^4(0)\bigg].
\ee 
In the simulations carried out here, $H/\tilde{\omega}_\vck\simeq 1/200$, making the $\mathcal{O}(H^4)$ corrections very small indeed. Specifically including a non-zero cosmological constant is straightforward, but we will not do so here.

Renormalization at NLO in 2PI in Minkowski space 
involves real-time solution of separate auxiliary equations and counterterms for each included diagram. This is because the 2PI diagrams resum self-insertions to all orders, and so the structure of divergences becomes more involved. Generalization to expanding backgrounds becomes even more complicated and is beyond the scope of this paper. We will simply note that although formally quadratically divergent, vacuum corrections to the mass from the Sunset diagram are numerically more than an order of magnitude smaller than the LO contribution, also at the largest coupling used below, $\lambda=6$ \cite{Arrizabalaga:2005tf}. At larger coupling this may no longer be the case, but then the coupling expansion should presumably be discarded altogether. Similar arguments apply to the vacuum contribution to the energy density.

This concludes our treatment of renormalization. Obvious refinements are possible, in particular if one is interested in subtle issues like particle creation from the vacuum, thermal (or warm) inflation. But for our present purposes, this will suffice.

% SUBSUBSECTION: INITIAL CONDITIONS

\subsection{Initial conditions\label{sec:initialconditions}}

We are interested in two types of initial conditions. The initial correlators are chosen gaussian, of the form
\be
\label{eq:initcorr}
\langle\varphi_\vck^{\dagger}\varphi_\vck\rangle=\frac{n_\vck^{\rm in}+1/2}{\Omega_\vck^{\rm in}},&\quad& \langle\partial_t\varphi_\vck^{\dagger}\partial_t\varphi_\vck\rangle=\left(n_\vck^{\rm in}+1/2\right)D_1(\mathcal{H})\Omega_\vck^{\rm in},\\
\langle\left[\partial_t\varphi_\vck^{\dagger},\varphi_\vck\right]\rangle=i,&\quad&\langle\left\{\partial_t\varphi_\vck^{\dagger},\varphi_\vck\right\}\rangle=D_2(\mathcal{H}),
\ee
in terms of a dispersion relation
\be
\Omega_\vck^{\rm in}=\bar{\Omega}_\vck\left(M\rightarrow M_{\rm gap}\right),
\ee
and a particle number in (approximate) equilibrium
\be
\label{eq:initeq}
n_\vck^{\rm in}=\left(e^{\frac{\omega_\vck^{\rm in}}{T}}-1\right)^{-1},~~~\textrm{Finite T},\qquad \omega_\vck^{\rm in}=\sqrt{\tilde{\bf k}^2+M_{\rm gap}^2},
\ee
and out of equilibrium
\be
n^{\rm in}_\vck=c,~~~ |\tilde{\bf k}|<c,~~~n^{\rm in}_\vck=0,~~~ |\tilde{\bf k}|>c,~~\textrm{Step}.
\ee
$T,c$ are free to be chosen. $T=0$ is the 
vacuum. 
The expansion is encoded in $\bar{\Omega}_\vck$ and the correction factors
\be
D_1(\mathcal{H})=1+\frac{\mathcal{H}^2}{(\omega_\vck^{{\rm in}})^2}\left(1+\left(\frac{M_{\rm gap}}{\omega_\vck^{\rm in}}\right)^2+\frac{1}{4}\left(\frac{M_{\rm gap}}{\omega_\vck^{\rm in}}\right)^4\right),\quad D_2(\mathcal{H})=-\frac{\mathcal{H}}{\omega^{\rm in}_\vck}\left(1+\frac{1}{2}\left(\frac{M_{\rm gap}}{\omega_\vck^{\rm in}}\right)^2\right).
\ee
$M^2_{\rm gap}$ is found by solving the 1-loop gap equation for the relevant initial condition,
\be
M^2_{\rm gap}=m^2_b+\frac{\lambda}{2}\left(\phi^2(0)+F_0\right),\quad F_0=\int_\vck \frac{n_\vck^{\rm in}+1/2}{\Omega_\vck^{\rm in}}.
\ee
The ``Step'' represents some generic out-of-equilibrium initial state, which is vacuum for large $k$. More elaborate choices are of course possible. ``Finite T'' is strictly speaking only equilibrium in the Hartree approximation at $H=0$ and at the initial time.

% SUBSUBSECTION: OBSERVABLES

\subsection{Observables\label{sec:observables}}

We monitor the global quantities physical time (\ref{eq:phystime}), scale factor $a(t)$, the Hubble rate, the mean field $\bar\phi(t)/M_{\rm pl}$, renormalized local correlator 
\be
\tilde{F}_{\rm ren}(t,t,{\bf x=0}) =\tilde{F}(t,t,{\bf x=0})-\delta\tilde{F}_{\rm vac}(t,t,{\bf x=0}),
\ee 
and renormalized energy density $\langle T^{00}(t)\rangle_{\rm ren}$. 

% SUBSECTION: PARTICLE NUMBERS *******************************************************************

\subsubsection{Particle numbers\label{sec:particlenumbers}}

For consistency with the renormalization and initialization prescriptions above, we should define particle number relative to the vacuum defined by $\bar{\Omega}_\vck$ (\ref{eq:omegadef}), to the same order in $H$. We use
\be
a^3(t)\tilde{n}_\vck^{\rm vac}(t)=a^3(t)\left(\frac{1}{2\bar{\Omega}_\vck(t)}\langle\dot{\varphi}_\vck(t)\dot{\varphi}_\vck(t)\rangle^{\rm vac}+\frac{\bar{\Omega}_\vck(t)}{2}\langle\varphi_\vck(t)\varphi_\vck(t)\rangle^{\rm vac}\right)-\frac{1}{2}.
\ee
By inserting $\bar{\Omega}_\vck$, we find
\be
a^3(t)\tilde{n}_\vck^{\rm vac}(t)=\frac{\mathcal{H}^2}{4(a\tilde{\omega}_\vck)^2}\left(1+\left(\frac{aM}{a\tilde{\omega}_\vck}\right)^2+\frac{1}{4}\left(\frac{aM}{a\tilde{\omega}_\vck}\right)^4\right)+\mathcal{O}\left(\dot{a}^4\right).
\ee
As expected this is zero in Minkowski space. The particle number relative to our vacuum is therefore
\be
a^3(t)\tilde{n}_\vck(t)=a^3(t)\left(\frac{1}{2\bar{\Omega}_\vck(t)}\langle\dot{\varphi}_\vck(t)\dot{\varphi}_\vck(t)\rangle+\frac{\bar{\Omega}_\vck(t)}{2}\langle\varphi_\vck(t)\varphi_\vck(t)\rangle\right)-\frac{1}{2}-a^3(t)\tilde{n}_\vck^{\rm vac}(t).
\ee
Out of equilibrium, it is advantageous to use instead the ``self-consistent'' definitions \cite{Salle:2000hd,Aarts:2001qa,Arrizabalaga:2005tf}
\be
a^3(t)n_\vck(t)&=&a^3(t)\sqrt{\langle\dot{\varphi}_\vck(t)\dot{\varphi}_\vck(t)\rangle\langle\varphi_\vck(t)\varphi_\vck(t)\rangle}-\frac{1}{2}-a^3(t)\tilde{n}_\vck^{\rm vac}(t),\\
\omega_\vck^{\rm eff}&=&\sqrt{\langle\dot{\varphi}_\vck(t)\dot{\varphi}_\vck(t)\rangle/\langle\varphi_\vck(t)\varphi_\vck(t)\rangle}.
\ee
Way out of equilibrium, $\omega_\vck^{\rm eff}$ will look wild, but close enough to equilibrium the two definitions agree in Minkowski space. With expansion it is easy to see that in our vacuum $\tilde{n}_\vck(t)=n_\vck(t)+\mathcal{O}\left(\dot{a}^4\right)$, and that
\be
\omega_\vck^{\rm eff}=\bar{\Omega}_\vck\left(1+\frac{\mathcal{H}^2}{2(a\tilde{\omega}_\vck)^2}\left(1+\frac{(aM)^2}{(a\tilde{\omega}_\vck)^2}+\frac{1}{4}\frac{(aM)^4}{(a\tilde{\omega}_\vck)^4}\right)\right).
\ee
In practice, the two definitions of $n_\vck(t)$ were seen to agree.

In the context of thermalisation, we quote the effective temperature and chemical potential of a mode, based on assuming a Bose-Einstein thermal distribution asymptotically\footnote{We note that $a^{3}(t)n_\vck$ rather than $n_\vck(t)$ itself is the quantity that is expected to equilibrate to a Bose-Einstein distribution, as a result of defining  Fourier transforms in terms of co-moving $\vck$.},
\be
\label{eq:lognk}
a^{3}(t)n_\vck=\left(\exp\left(\frac{\omega_\vck^{\rm eff}-\mu_{\rm ch}^{\rm eff}}{T_\vck^{\rm eff}}\right)-1\right)^{-1}\rightarrow \ln\left(\frac{1}{a^{3}(t)n_\vck}+1\right)=\frac{\omega_\vck^{\rm eff}-\mu_{\rm ch}^{\rm eff}}{T_\vck^{\rm eff}}.
\label{eq:Teff}
\ee
When $T_\vck^{\rm eff}$ and $\mu_{\rm ch}^{\rm eff}$ are independent of $\vck$, we say that the system has equilibrated kinetically. In Minkowski space the timescale for this to happen for $\lambda=6$ (as used throughout most of this paper) is $\simeq 1000m^{-1}$ \cite{Arrizabalaga:2005tf}. Chemical equilibration is when this common $\mu^{\rm eff}_{\rm ch}$ is zero. Again in Minkowksi space, this is roughly an order of magnitude slower than kinetic equilibration.

% SECTION EXAMPLE APPLICATIONS

\section{Example applications\label{sec:examples}}

The setup introduced above can now be applied to systems and phenomena of interest. We use lattices of $32^3$ points, time-steps $\delta t=0.05$, lattice spacing $\delta x=1$. The masses are $m/\mpl=0$ (massless) $m/\mpl=0.00025-0.001$ (massive), and for the interacting case $\lambda=6$ to study thermalisation, $\lambda=0.1$ for the case of preheating. Thermal initial conditions have $T/\mpl=0.0005-0.002$ and out-of-equilibrium Step initial conditions have $c/m=5$. We use $M_{\rm pl}$ to set the scale of gravity and hence the expansion rate. Large $M_{\rm pl}$ means slower expansion. When $a(t)m\simeq 1$, one should start worrying whether the lattice is too coarse. When $a(t)m>2$, results can probably no longer be trusted. In the massive case, $a(0)m=0.2$.

From the point of view of a realistic cosmology, at least after inflation our choice of relative $M_{\rm pl}$ is much too small, and our coupling too large. In  $m^2\varphi^2$ inflation $m/M_{\rm pl}<10^{-5}$ and in $\lambda\varphi^4$ inflation $\lambda\simeq 10^{-14}$, to be consistent with measurements of the Cosmic Microwave Background. Typical reheating temperatures are or the order of $10^{-8}M_{\rm pl}$. Therefore the parameters here over-emphasize expansion effects compared to most post-inflationary cosmological phenomena. Inflation itself is an exception, since the relatively much larger couplings used here (in the preheating application, $0.1$ instead of $10^{-14}$) may make larger interaction rates compensate for the larger expansion.

It may also be worth pointing out that as $H< 10^{-3}$ (see below), all the modes under consideration are sub-horizon, 
\be
k_{\rm min}\simeq 0.2/a(t)\gg H,
\ee
since we only allow $a(t)<10$.

% SUBSECTION: FREE FIELD AT FINITE TEMPERATURE

\subsection{Free field at finite temperature\label{sec:free_thermal}}

\begin{figure}[t!]
\begin{center}
\epsfig{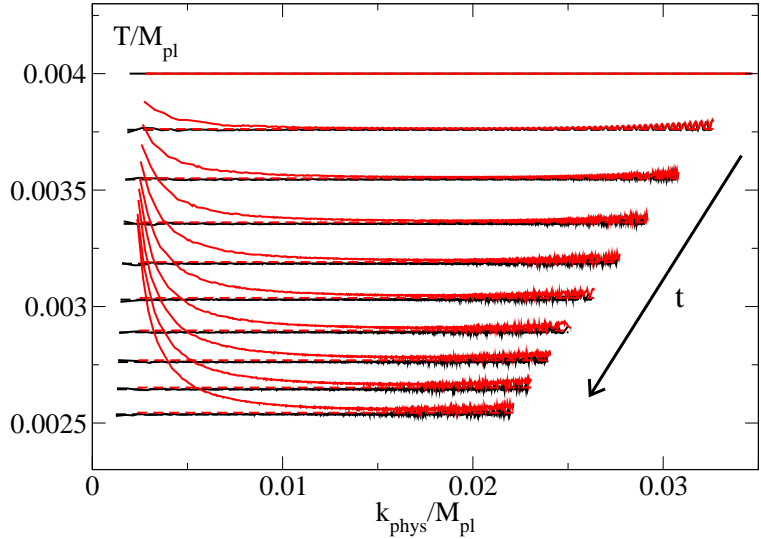}
\caption{The effective temperatures $T_\vck^{\rm eff}$ for all modes $k$ in a free simulation starting from equilibrium. In the massles case (black), the modes stay thermal with a temperature dropping as $1/a$. In the massive case (red) this is only true in the UV. The range of physical momenta redshift towards the IR. Overlaid are (dashed) lines denoting $T(t=0)/a(t)$ in the two cases. }
\label{fig:freeT2}
\end{center}
\end{figure}

The simplest case is a free field initially in the vacuum. But since we have effectively chosen zero cosmological constant as our renormalization condition, the Universe is just static. Still, it is a useful test of the numerics and the renormalization.

A slightly more interesting test case is to initialize a free field in a thermal state, $\lambda=0$, $T/M_{\rm pl}=0.04$, $m/M_{\rm pl}=0.02;\,0$, shown in Figure \ref{fig:freeT2}. The effect of the expansion is to redshift the momenta $\vck_{\rm phys}(t)=\tilde{\vck}=\vck/a(t)$, but the particle number as a function of co-moving momentum $a^3n_\vck(t)$ is constant. If the field is massless, we have
\be
a^3n_\vck(t)=f\left(\omega_\vck/T \right)=f\left(|\vck|/T \right),
\label{eq:fscaling}
\ee
with in our case $f(x)=(\exp(x)-1)^{-1}$. If we identify the constant $T$ as the temperature, then $f(|\vck|/T)$ is constant if $T(t)=T(0)/a(t)$. Hence for the massless case, $T^{\rm eff}_\vck$ (\ref{eq:Teff}) should be independent of $\vck$ and decrease as $1/a$. These are the black lines of Figure \ref{fig:freeT2}. In the massive case, we no longer have (\ref{eq:fscaling}), since $\tilde{\omega}_\vck=\sqrt{m^2+|\vck|^2/a^2}$. For $|\vck|\gg m$ the argument still holds, but for small $|\vck|$ we have deviation from $|\vck|$-independence. This is shown by red lines. We may choose to interpret the deviation as a chemical potential, and it is easy to see\footnote{Use that $\tilde{\omega}_\vck\simeq m+k^2/(2a^2m)$. Then for $[\tilde{\omega}_\vck-\mu_{\rm ch}^{\rm eff}(t)]/T(t)$ to be constant, $\mu_{\rm ch}^{\rm eff}(t)=m(1-1/a^2(t))$.} that for $|\vck|\ll m$, $T\propto 1/a^2(t)$ and $\mu_{\rm ch}^{\rm eff}/m=1-1/a^2(t)$. We stress that at this point the chemical potential is only meant to express a relative over-abundance of particles in the low momentum modes.

As a consequence of redshift only, we therefore expect that a non-interacting massive field will have an effective chemical potential going asymptotically to $\mu_{\rm ch}^{\rm eff}/m=1$. When including interactions, these will drive the chemical potential towards zero, but only if they are fast enough on the time-scale of the expansion.

We can trust the simulation as long as the high $|\vck|$ modes stay in the vacuum, with no significant thermal population. This is automatic in the free field case, but when including interactions, high $|\vck|$ modes will be excited.

\subsection{Free mean field and vacuum modes\label{sec:free_mean}}

\begin{figure}[t!]
\begin{center}
\epsfig{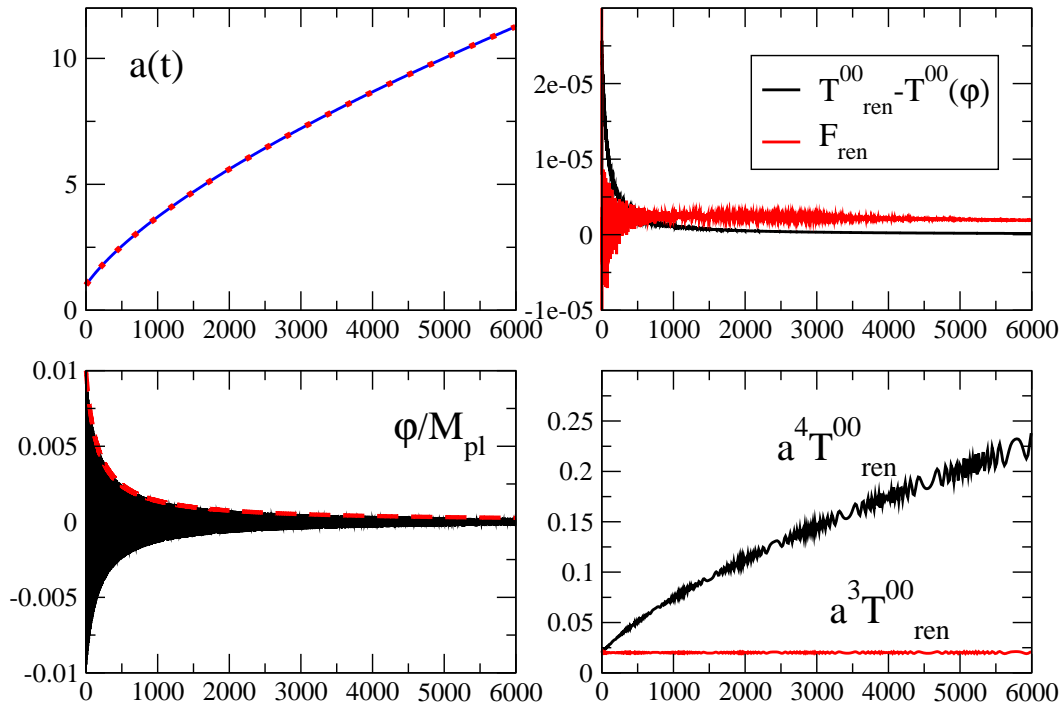}
\caption{The scale factor (upper left), mean field (lower left), renormalized correlator and energy density (right) in the presence of an oscillating mean field. Both correlator and energy density are well renormalized, and the energy density behaves as matter, $\propto a^{-3}$. The scale factor evolves as matter domination $a=(1+t/t_0)^{2/3}$ (overlaid), and the mean field amplitude decreases as $1/t$ (overlaid).}
\label{fig:freeMF}
\end{center}
\end{figure}

Another interesting check of the numerics is to initialize the mean field away from zero and let it oscillate freely, without interacting with the modes. These are initialized in the vacuum. It is easy to see that an oscillating, homogeneous mean field behaves as matter, with zero pressure, $\langle T^{00}\rangle_{\rm ren}\propto a^{-3}$ and $a(t)\propto t^{2/3}$. Also, the mean field is expected to have the form
\be
\label{eq:mfosc}
\bar{\phi}(t)=\frac{\phi_0}{1+t/t_0}\cos(mt).
\ee
This supposes that the vacuum has been correctly renormalized and does not contribute to the energy density. In principle, fast expansion could induce particle production (an $\mathcal{O}(H^4)$ effect), but this will be very small indeed.

In Figure \ref{fig:freeMF} we show the scale factor (upper left), the mean field (bottom left) and the renormalized energy density (bottom right). They all scale with $a(t)$ as expected. In the top right frame, we show the renormalized energy density after the mean field contribution is subtracted, and the renormalized equal time correlator. Both are renormalized to $10^{-5}$ (here in units where the un-renormalized quantity is $\mathcal{O}(1)$).

% SUBSECTION: INTERACTING FIELD AT FINITE TEMPERATURE

\subsection{Interacting field at finite temperature\label{sec:interacting_thermal}}

\begin{figure}[t!]
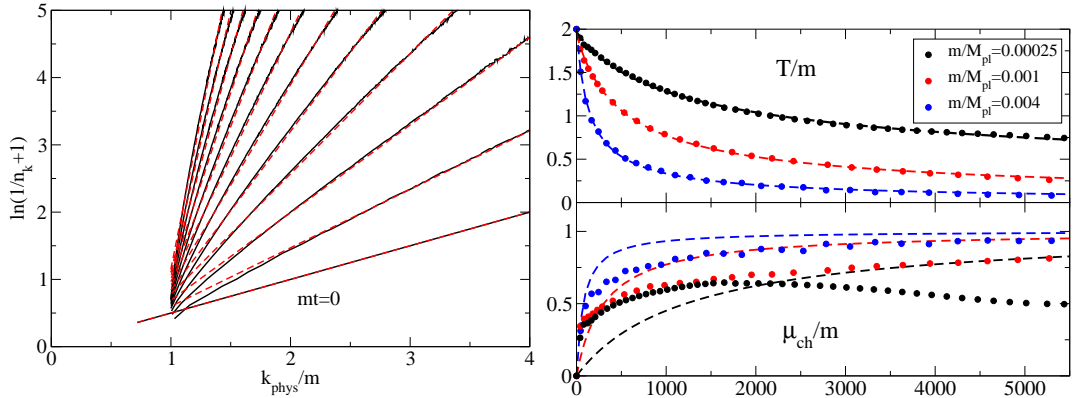

\begin{center}
\epsfig{file=./pictures/spectrum_T2int_ex.eps,width=7cm,clip}
\epsfig{file=./pictures/muT_allT2.eps,width=7cm,clip}
\caption{Left: The mode spectrum in time, starting from a thermal initial condition. The observable is chosen so that it is a straight line in kinetic equilibrium (\ref{eq:lognk}), and effective temperature and chemical potentials can be found from a fit (overlaid). Right: Effective temperature and chemical potential for a fast ($T^{\rm init}/M_{\rm pl}=0.008$), slow ($T^{\rm init}/M_{\rm pl}=0.002$) and very slow expansion ($T^{\rm init}/M_{\rm pl}=0.0005$). The temperature drops as matter domination, and the chemical potential becomes larger for faster expansion, when interactions have a harder time keeping up. Dashed lines represent the free field behaviour $\mu_{\rm ch}/m=1-1/a^2(t)$.}
\label{fig:T2modes}
\end{center}
\end{figure}

Interactions alter the dynamics and the thermal state is only approximately Bose-Einstein, with an additional thermal mass component. A fair approximation to an initial thermal state is to solve for the thermal mass at LO, and use this to generate a Bose-Einstein distribution (\ref{eq:initeq}). Because we evolve with NLO equations of motion, the initial condition is not exactly the equilibrium one. Also, since we include expansion, the system can anyway only be in approximate equilibrium. We use $\lambda=6$, $T_{\rm init}/M_{\rm pl}=0.0005,0.002,0.008$, $T_{\rm init}/m=2$.

Figure \ref{fig:T2modes} (left) shows $\ln(1/n_\vck+1)$ (\ref{eq:lognk}) in time. 
%Kinetic equilibrium is reasonably maintained throughout. 
We see that interactions partially compensate for the chemical potential $\mu_{\rm ch}/m=1-1/a^2(t)$, although kinetic equilibration is not complete. We may tentatively extract effective temperatures and chemical potentials from fits to the spectrum, shown in the right-hand panel. In all cases, temperature drops roughly as matter domination. As for the non-interacting case, the chemical potential rises and goes asymptotically to a finite value. Clearly, interactions are not fast enough to uphold or restore chemical equilibrium. In the fast expansion case, although the rise is slower than $m(1-1/a^2(t))$ (the initial mass is not the zero-temperature one), asymptotically it goes to $m$, suggesting that interactions are ineffective. In the slow case, the asymptotic value is about $3/4$ as large, as a result of interactions. In the very slow case, a slight decrease can be seen at late times. Still, in all cases, a relic abundance of particles freezes in.

A common criterion for staying thermal is for some interaction rate $\Gamma$ to dominate the Hubble rate. Such a rate could be provided by the Sunset diagram damping rate \cite{Wang:1995qf}. At large temperature $T\gg m$\footnote{We note that our system is probably not in this asymptotic regime, and not at small coupling, and so a more complicated expression may be required for detailed estimates. For an order of magnitude estimate, however, this will do (see also \cite{Arrizabalaga:2005tf}).},
\be
\label{eq:sunsetdamping}
\Gamma(T,\lambda)\simeq\frac{\lambda^{3/2} T}{50\pi^2}.
\ee
The Hubble rate is given in terms of the energy density
\be
H^2 = \frac{1}{3M^2_{\rm pl}}\frac{\pi^2}{30}T^4,
\ee
and so with our parameters
\be
\label{eq:HGratio}
\frac{\Gamma}{H}\simeq 0.09\frac{M_{\rm pl}}{T}\gg 1,
\ee
suggesting that the system should be able to stay thermal. Even more so as temperature drops, although when $m\simeq T$ (\ref{eq:HGratio}) no longer applies. The ``scattering time scale'' $\Gamma^{-1}$ is $mt_{\rm scat.}\simeq \frac{M}{15T}\simeq 17$. Chemical equilibration is known to require of order $500\, t_{\rm scat.}$ at this coupling in Minkowski space \cite{Arrizabalaga:2005tf}. Therefore $\Gamma/H\gg 1$ is not sufficient for chemical equilibration, although as we will see below, kinetic equilibration does take place.

The 2PI-resummed Sunset diagram includes 2-to-2 scattering as well as off-shell 1-to-3 scattering. The damping rate (\ref{eq:sunsetdamping}) is dominated by the former, which leads to kinetic but not chemical equilibration, as particle number is conserved. The latter off-shell process does change particle number, but is a higher order effect included in the diagram through the 2PI resummation of many perturbative diagrams. It is therefore no surprise that chemical equilibration happens on a much longer timescale, being (naively) suppressed by additional powers of the coupling. This highlights the possible shortcomings of criteria like (\ref{eq:HGratio}) to establish thermalization.

% SUBSECTION: THERMALISATION WITH EXPANSION

\subsection{Thermalisation with expansion\label{sec:thermalisation}}

\begin{figure}[t!]
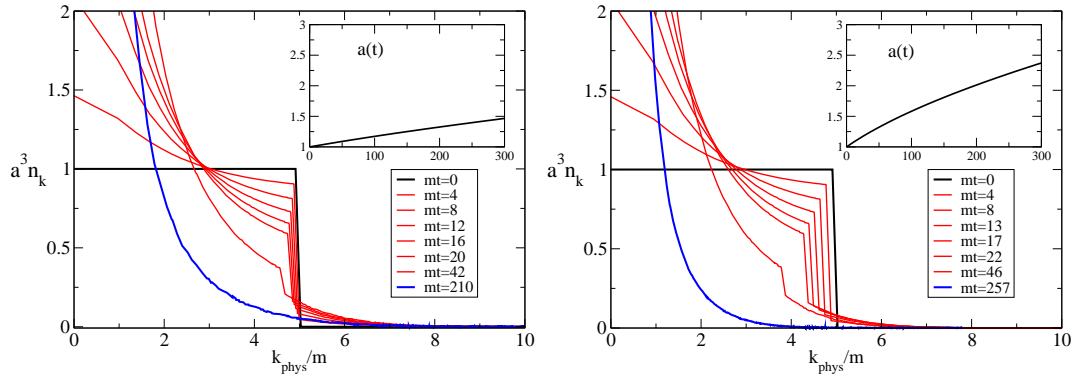

\begin{center}
\epsfig{file=./pictures/tsuspectrum200_1.eps,width=7cm,clip}
\epsfig{file=./pictures/tsuspectrum50_1.eps,width=7cm,clip}
\caption{The spectrum $a^3n_\vck$ vs. $k$ when starting from a Step initial condition. Again, left is for $T^{\rm init}/M_{\rm pl}=0.002$, right for $T^{\rm init}/M_{\rm pl}=0.008$. Kinetic equilibration happens on timescales of order $mt=200$ in both cases. The resulting temperature is clearly smaller for fast expansion.}
\label{fig:tsuspectrum}
\end{center}
\end{figure}

In order to study kinetic equilibration, and its dependence on expansion rate, we start the system way out of equilibrium using the Step initial condition. Figure \ref{fig:tsuspectrum} shows the evolution of the spectrum at early times, for a slow expansion ($m/M_{\rm pl}=0.001$, left) and a fast one ($m/M_{\rm pl}=0.004$, right). The Step is smeared out into a smooth spectrum on a timescale of $mt\simeq 200$, while the modes are redshifted towards the infrared. In this particular case, redshift may help equilibration, as evolution from the Step to equilibrium involves transfer of power to smaller $|\vck|$ modes. For even faster expansion, this redshift may ``overtake'' the equilibrium distribution, and intermediate-$|\vck|$ modes will have to be re-populated through scattering.

\begin{figure}
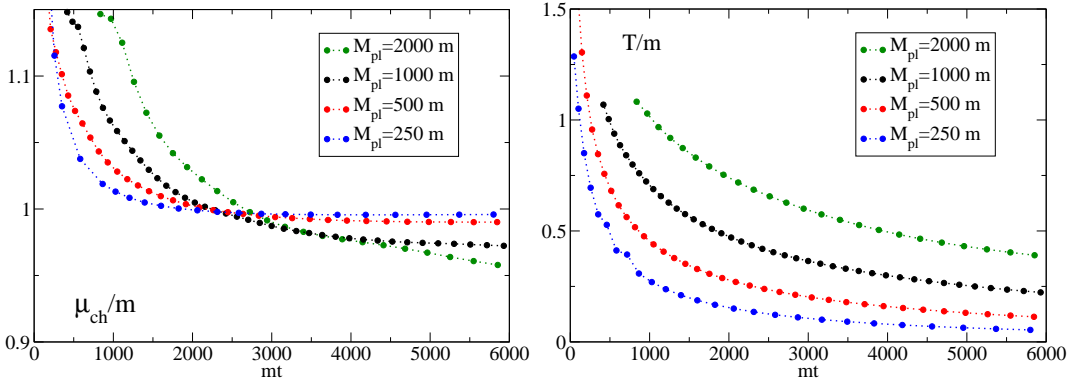

\begin{center}
\epsfig{file=./pictures/tsu_mu.eps,width=7cm,clip}
\epsfig{file=./pictures/tsu_T.eps,width=7cm,clip}
\caption{Effective chemical potentials (left) and temperatures (right) in time for different expansion rates. Only after some time are the spectra sufficiently straight to allow fits to determine $T^{\rm eff}$ and $\mu_{\rm ch}^{\rm eff}$. The asymptotic chemical potential increases with increasing expansion rate, interactions become more inefficient and more particles freeze in. Final temperature decreases with expansion rate as a power law, $t^{-(0.4-0.8)}$.}
\label{fig:tsumuT}
\end{center}
\end{figure}

Once the spectrum is kinetically equilibrated, evolution is much slower, and we can again quantify the spectrum by the effective temperature and chemical potential. This is shown in Figure \ref{fig:tsumuT}, with the $T^{\rm eff}$ on the left and the $\mu_{\rm chem}^{\rm eff}$ on the right. Temperature drops as $t^{-\alpha}$ with $\alpha$ between $0.4$ and $0.8$, while the chemical potential again asymptotes to a finite value. Faster expansion (smaller $M_{\rm pl}$) again results in a larger asymptotic value. 

We are therefore in a regime where interactions are strong enough to equilibrate an initial out-of-equilibrium condition into a Bose-Einstein-like thermal state, but not strong enough to get rid of the chemical potential.
As in the section above, we conclude that the $\Gamma/H\gg 1$ criterion applies to restoration and/or maintaining of kinetic equilibrium only.

% SUBSECTION: WITH A MEAN FIELD: REHEATING

\subsection{Preheating after inflation\label{sec:reheating}}

At the end of cosmological inflation, the inflaton mean field leaves the slow-rolling stage and begins oscillating around a minimum of its potential. Through interaction with other fields and/or its own field modes, energy is transferred into particle excitations of these fields. In some cases, resonant particle creation can take place, known as preheating \cite{Kofman:1994rk}. This usually lasts for a few periods of the inflaton oscillation, during which some fraction of the energy is transferred, after which normal perturbative decay and reheating transfers the rest.

\begin{figure}[t!]
\begin{center}
\epsfig{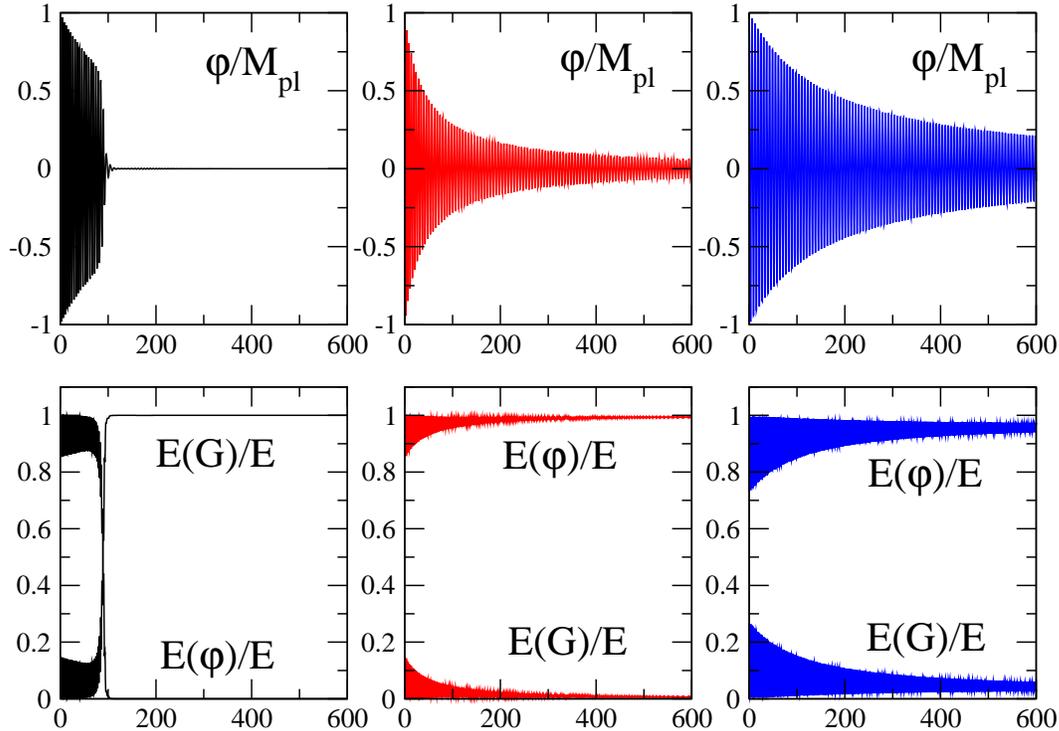}
\caption{An oscillating mean field coupled to its modes results in reheating and possibly resonant preheating. Shown is the mean field (upper)
and the energy components (lower)
for three cases: Large field, slow expansion (left); large field, fast expansion (middle), small field, slow expansion (right). For the first case, around $mt=100$ the mean field dumps all its energy into particles through resonant preheating.}
\label{fig:reheating}
\end{center}
\end{figure}

Let us assume for simplicity that the mean field oscillates harmonically as (\ref{eq:mfosc}), including the self-interaction (which induces an-harmonic oscillations) only through an altered frequency $m_{\rm osc}$
\be
\label{eq:phiosc}
\bar{\phi}(t)=\phi_0(t)\cos(m_{\rm osc}t).
\ee
Then each mode function $\varphi_\vck(t)$ obeys approximately
\be
\ddot{\varphi}_\vck+3H(t)\dot{\varphi}_\vck+\left(\frac{k^2}{a^2(t)}+m^2+\frac{\lambda}{2}\phi_0^2(t)\cos^2(m_{\rm osc}t)\right)\varphi_\vck=0.
\ee
Again approximately, a mode $\vck$ will be in resonance if
\be
\label{eq:Ak}
A_\vck\equiv\frac{\vck^2}{a^2(t)m_{\rm osc}^2}+\frac{m^2}{m_{\rm osc}^2}+\frac{\lambda\phi_0^2(t)}{8m_{\rm osc}^2}=l^2,
\ee
with $l$ an integer, with the strongest resonance for $A_\vck=1$. In the resonance, particle number grows exponentially. Because $a(t)$ and the amplitude $\phi_0(t)$ decrease in time, in an expanding background, modes will move in and out of resonance. 

Figure \ref{fig:reheating}, upper panels, shows the mean field for three different cases, normalized to their initial amplitude: $m/M_{\rm pl}=0.001$, $\bar\varphi(0)/M_{\rm pl}=0.01$ (black); $m/M_{\rm pl}=0.004$, $\bar\varphi(0)/M_{\rm pl}=0.04$ (red); and $m/M_{\rm pl}=0.004$, $\bar\varphi(0)/M_{\rm pl}=0.01$ (red). Since energy density is dominated by the mean field, the expansion rates are the same initially in the first and second case, faster in the third. But $M_{\rm pl}$ is different in lattice units and in units of $m$ in the first case. Around time $mt=100$ the mean field oscillation suddenly collapses for the slow expansion, small $m/M_{\rm pl}$, case only. In the lower panels, we see that at the same time, energy is transferred from the mean field component ($E(\phi)$) to the mode component ($E(G)$) (normalised to the total energy). We conclude that we have preheating only in this one case. 

\begin{figure}[t!]
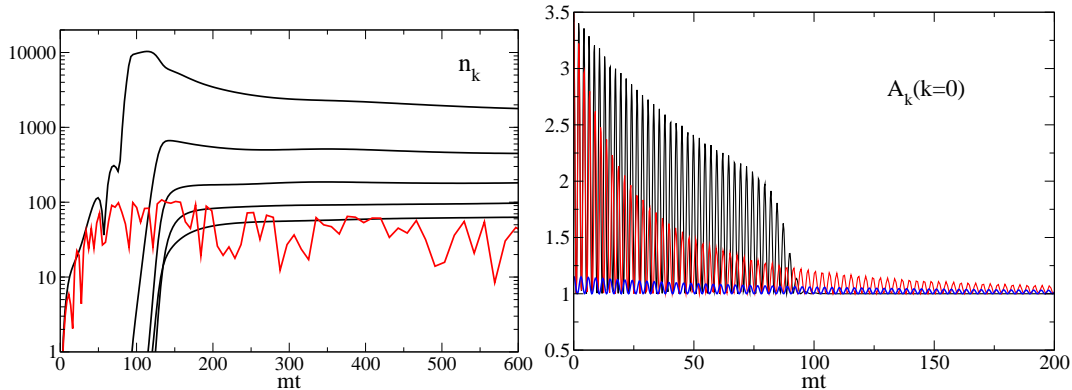

\begin{center}
\epsfig{file=./pictures/MF_global5.eps,width=7cm,clip}
\epsfig{file=./pictures/MF_global4.eps,width=7cm,clip}
\caption{Left: Particle numbers of the 5 lowest $|\vck|$ modes. Only in the large field cases is there a resonance, and only in the slowly expanding cases (black) is the resonance long enough to get large particle numbers. With fast expansion (red) only the zero mode grows briefly. Right: $A_\vck$ for the zero modes (\ref{eq:Ak}), with the replacement $\bar\phi_0\rightarrow \bar\phi(t)$. Hence $A_\vck$ proper is the envelope of the oscillating curves. }
\label{fig:reheating2}
\end{center}
\end{figure}

Indeed, in figure \ref{fig:reheating2} (left), we show particle numbers of the 5 lowest momentum modes; these grow exponentially, with the first (zero) mode leading the way early on reaching $n_\vck\simeq 10000$. But also in the fast expanding case (red), the zero mode grows. In that case the resonance ends rather early on, particle number is only order 100, and the non-zero modes are not excited. In the right panel, $A_\vck$ for the zero mode is the envelope of the oscillating curves, shown for the three cases. The overall normalisation will depend on $m_{\rm osc}$, and apparently the resonance band is somewhere above $A_\vck=1$. The qualitative picture is fairly clear. The blue curve is below the resonance, and no preheating occurs. The (amplitude of the) black curve goes through the resonance band long enough for a significant amplification of the zero mode. This eventually excites the non-zero, non-resonating modes to complete reheating altogether. For faster expansion (red), the mean field amplitude decays very fast, and apparently the zero mode is only briefly in resonance, but does not grow sufficiently to trigger substantial preheating.

Clearly a more detailed study is required, for finer lattices where not only the zero mode resonates, and taking into account that the mean field oscillation is anharmonic. Also, in most realistic models of inflation, the mean field starts out $\mathcal{O}(M_{\rm pl})$. Finally, a realistic (self-)coupling would be many orders of magnitude smaller than the one employed here, $\lambda=0.1$.  Such a study is beyond the scope of the present work. Still, resonant preheating is within the range of application of the present formalism, and outcomes depend strongly on the expansion rate.

Parametric resonance was studied using 2PI methods in \cite{Berges:2002cz}, in Minkowski space. Particle numbers grow exponentially to $n_\vck\simeq 1/\lambda$, and so in principle a 2PI coupling expansion like the one used here may not be applicable. Indeed, for small coupling, numerical instabilities are experienced. In \cite{Berges:2002cz} a $1/N$ expansion in the number of fields is used, making the dynamics more stable. Using the procedure outlined here, this can be generalized to FRW spaces. 

% SECTION: CONCLUSION

\section{Conclusion\label{sec:conclusion}}

We have seen that extending the 2PI formalism of out-of-equilibrium quantum fields to expanding backgrounds amounts to introducing a time dependent mass
$m^2\rightarrow a^2m^2-a''/a$ in the conformal time, rescaled field, equations of motion (10-19). Having solved these equations on the lattice, observables are translated back to physical fields and physical time. The scale factor is derived from the semi-classical Friedmann equation involving the renormalized energy $\langle T^{00}\rangle_{\rm ren}$. We have here opted for an approximate renormalization strategy, where counterterms for the mass and the energy density are calculated in the LO/Hartree approximation in terms of a particular vacuum, the adiabatic free-field solution to second order, both in WKB and in $\dot{a}$. We argued that going beyond this order is possible, although only really necessary when aiming at taking the continuum limit or using very large expansion rates. A fully 2PI renormalization beyond LO is much harder.

On a finite comoving lattice, modes are redshifted towards the IR. Therefore the number of e-folds of expansion available is limited; a simulation can only be trusted as long as there is a range of UV modes that stay in the vacuum. Otherwise cut-off effects will influence the physics and presumably the renormalization. One way of quantifying this is for the time-dependent mass $am$  to stay less than unity. In the ``conformal'' case $m=0$ another mass scale (temperature, initial mean field) will play a similar role. 

At the end of the day this is a practical question of computer capacity. 2PI simulation are memory intensive in that the memory of past time-steps must be saved to generate the self-energy kernels (right hand sides of (10-12)). On the other hand, no statistical averaging is necessary as the solutions to the equation are the full correlators. The total simulations performed here amount to about 5000 CPU hours. 

Possible applications range over all the topics already studied in Minkowski space: Thermalisation \cite{Berges:2000ur,Berges:2002wr,Juchem:2003bi,Berges:2004ce,Arrizabalaga:2005tf,Lindner:2005kv}, which enters in early Universe physics as well as heavy ion collisions\footnote{In heavy ion collisions the expansion is somewhat different from FRW.}; reheating and preheating, both the resonant variety \cite{Berges:2002cz} and tachyonic preheating \cite{Felder:2000hj,Arrizabalaga:2004iw}. In the latter cases, it may be prudent to use a diagram expansion in $1/N$ rather than $\lambda$, to be sure corrections are under control.

In the present paper we have made test-runs of many of these cases and pointed out the main effects of the cosmological expansion. All of these applications deserve further scrutiny, also of combined effects of smaller/larger couplings, expansion rate, temperature and masses. A study of the inflationary regime was not attempted, mainly because it would require very large lattices (lots of expansion) and/or $m=0$ which is a very special case.

One interesting result is the difference between the kinetic and chemical equilibration timescales, and the possibility of having one but not the other happen. Clearly, one must be careful when assuming instantaneous thermalisation, as is sometimes done when considering reheating, preheating and phase transitions in the early Universe. Using a criterion like $\Gamma/H\gg 1$ to ensure instant thermalisation presumes careful consideration of which $\Gamma$ is the relevant one.

We were also able to confirm that the amount of resonant preheating depends sensitively on the rate of expansion. This is because field modes are redhifted in and out of resonance bands. This shortens the resonance time, making even exponential growth much less effective.

The quantum 2PI equations have a classical counterpart, i.e. 2PI-truncated equations for classical correlators, reproducing classical dynamics \cite{Aarts:2001yn}. The relation between the two amounts to neglecting terms like $\rho^2$ compared to terms like $F^2$ in the self-energies (14-16), and ignoring renormalization. Although the full classical approximation is in principle exact, it relies on  statistical averaging over initial conditions. Classical 2PI has no statistical errors, but diagram expansion truncation introduces a different type of approximation. As such, it contributes an alternative way of doing classical simulations. 

Extension to more complicated models than a single self-interacting scalar is straightforward. Possible applications include reheating and preheating with multiple fields (including fermions \cite{Berges:2002wr}), departure from and return to equilibrium for systems with heavy particles decaying into light ones. In the context of multi-field preheating this may allow to calculate non-gaussian signatures in the CMB \cite{Chambers:2007se,Chambers:2008gu}. 

In conclusions, we believe that given sufficient numerical capacity, and observing certain simple rules, the 2PI formalism for out-of-equilibrium fields provides a convenient tool for quantitative calculations in cosmology.

% ACKNOWLEDGEMENTS

\vspace*{0.5cm}
\noindent
{\bf Acknowledgments.}
It is a pleasure to thank Gert Aarts, Kari Rummukainen and Szabolcs Borsanyi for many comments, and Arttu Rajantie and Jan Smit for enlightening discussions and work on related topics. I am indebted to Julien Serreau for a host of useful suggestions and for pointing out a couple of errors. The numerical work was conducted on the Murska cluster at the Finnish center for computational sciences, CSC. This work was supported by Academy of Finland Grant 114371.

% SECTION: ADIABATIC REGULARISATION ************************************************

% SUBSUBSECTION: STRUCTURE OF DIVERGENCES
\appendix

\section{Structure of divergences\label{sec:structurediv}}

Here, we present a detailed exposition of the cancellation of divergences in the renormalization procedure presented in the main text.

We begin in Minkowski space, at 2PI-LO. At this order, the equation of motion for the propagator modes reads,
\be
\left[\partial_t^2+\tilde\omega^2_\vck(t)\right]\tilde{F}(t,t,\vck)=0,
\ee
with
\be
\tilde\omega_\vck^2(t)=\vck^2+m^2_b+\frac{\lambda}{2}\tilde{F}(t,t,{\bf x=0})+\frac{\lambda}{2}\bar{\varphi}^2(t).
\ee
Away from the vacuum, and for the moment neglecting time derivatives of the mass, write\footnote{Note that although we have an interacting theory, at 2PI-LO the field can still be written in terms of mode functions, and are therefore completely described in terms of a particle number and a dispersion relation with an effective time-dependent mass.}
\be
\tilde{F}(t,t,{\bf x=0})=F_{\rm vac}+F_{\rm res}(t)=\int_\vck\frac{n_\vck(t)+1/2}{\tilde\omega_\vck^2(t)}.
\ee
In our prescription, we choose
\be
F_{\rm vac}=\int_\vck\frac{1/2}{\sqrt{\vck^2+m^2+\frac{\lambda}{2}\bar{\varphi}_0^2}},
\ee
and renormalize the mass by
\be
m_b^2+\frac{\lambda}{2}F_{\rm vac}=m^2.
\ee
We will also need
\be
F_0=\int_\vck\frac{1}{2}\sqrt{\vck^2+m^2+\frac{\lambda}{2}\bar{\varphi}_0^2}.
\ee
Then 
\be
\tilde\omega_\vck^2(t)=\vck^2+m^2+\frac{\lambda}{2}F_{\rm res}(t)+\frac{\lambda}{2}\bar{\varphi}^2(t).
\ee
The energy density can be written
\be
\langle T^{00}(t)\rangle_{\rm ren}&=&\frac{1}{2}\int_\vck\left(\partial_t\partial_{t'}+\tilde\omega_\vck^2(t)\right)\tilde{F}(t,t',{\bf k})_{t=t'}\nonumber\\&&-\frac{\lambda}{8}\tilde{F}(t,t,{\bf x=0})^2+\frac{m_b^2}{2}\bar{\varphi}^2(t)+\frac{\lambda}{24}\bar{\varphi}^4(t)+\delta T^{00}.
\ee
We can rewrite this as
\be
\label{eq:en4}
\langle T^{00}(t)\rangle_{\rm ren}&=&\frac{1}{2}m^2\bar{\varphi}^2(t)+\frac{\lambda}{24}\bar{\varphi}^4(t)-\frac{\lambda}{8}F_{\rm res}^2\nonumber\\
&&\int_\vck(n_\vck+1/2)\tilde\omega_\vck(t) -\frac{\lambda}{8}F_{\rm vac}^2+\delta T^{00}-\frac{\lambda}{4}F_{\rm vac}\left(F_{\rm res}+\bar{\varphi}^2(t)\right)
\ee
The first line has at most logarithmic divergences. The second line quartic, quadratic and logarithmic. Let us assume that the particle number $n_{\vck}$ decays faster than $k^{-4}$, in which case the divergencies result only from
\be
\int_{\vck}\frac{\tilde\omega_\vck(t)}{2}=F_0+\frac{\lambda}{4}\left(\bar{\varphi}^2(t)-\bar{\varphi}_0^2+F_{\rm res}(t)\right)F_{\rm vac}+\mathcal{O}\left({\rm logs}\right).
\ee
The middle term almost precisely cancels the last term in (\ref{eq:en4}), leaving only a divergent constant. Therefore, by choosing
\be
\delta T^{00}=\frac{\lambda}{8}F_{\rm vac}^2+\frac{\lambda}{4}\bar{\varphi}_0^2F_{\rm vac}-F_0,
\ee
all the remaining quartic and quadratic divergences are cancelled.

Let us now consider what happens when including time derivatives of the mass. We can consider this in the WKB approximation, in a similar way to the main text, but with
\be
\tilde\omega_\vck^2=k^2+M^2(t),\qquad M^2=m_b^2+\frac{\lambda}{2}\left(\bar{\varphi}^2+F_{\rm vac}+F_{\rm res}(t)\right).
\ee
The ansatz is
\be
\varphi_\vck(t)=a_\vck f_\vck(t)+a_\vck^\dagger f_\vck^{*}(t),\quad f_\vck(t)=\frac{1}{\sqrt{2\Omega_\vck(t)}}e^{-i\int^t\Omega_\vck(t')dt'}.
\ee
As above, we find that
\be
\Omega_\vck^2(t)=\tilde\omega_\vck^2(t)\left[1-\frac{1}{2}\frac{\ddot{\omega}_\vck}{\omega_\vck^3}+\frac{3}{4}\left(\frac{\dot\omega_\vck}{\omega_\vck^2}\right)^2\right].
\ee
with 
\be
\left(\frac{\dot{\omega}_\vck}{\omega_\vck^2}\right)_{\rm Minkowski}=\frac{M\dot{M}}{\omega_\vck^3}\qquad \left(\frac{\ddot{\omega}_\vck}{\omega_\vck^3}\right)_{\rm Minkowski}=\ldots.
\ee
We can compare this to the FRW case discussed above, where we have
\be
\frac{\dot{\omega}_\vck}{\omega_\vck^2}=-\frac{H}{\omega_\vck}\left(1-\frac{M^2}{\omega^2_\vck}\right).
\ee
When we calculate the energy density, we have for instance
\be
\int_\vck\partial_t\partial_{t'}F(t,t',{\bf x=0})_{t=t'}
&=&
\int_\vck\frac{\Omega_\vck}{2}\left(1+\frac{1}{4}\left(\frac{\dot{\Omega}_\vck}{\Omega_\vck^2}\right)^2\right)\nonumber\\
&=&
\int_\vck\frac{\omega_\vck}{2}\left(1-\frac{1}{4}\frac{\ddot{\omega}_\vck}{\omega^3_\vck}+\frac{5}{8}\left(\frac{\dot{\omega}}{\omega^2_\vck}\right)^2\right).
\ee
The leading term in time derivatives (no derivatives) is the usual quartically divergent term. In the FRW case, the next order is of the form
\be
\propto \int_\vck\frac{H^2\textrm{ or }\dot{H}}{\omega_\vck},
\ee
which is quadratically divergent. However, from the time derivative of the mass, we get terms of the form
\be
\propto \int_\vck\frac{\omega_\vck}{2}\frac{\dot{M}^2\textrm{ or }M\ddot{M}}{\omega_\vck^4},
\ee
which are logarithmically divergent as well as finite terms, and in FRW we in addition get mixed terms like
\be
\propto \int_\vck\frac{\omega_\vck}{2}\frac{HM\dot{M}}{\omega_\vck^4},
\ee
which are also logarithmically divergent. 

Therefore, since we choose to ignore logarithmic divergences in the energy, no counterterms proportional to the derivative of the mass are necessary.

% THE BIBLIOGRAPHY ************************************************************

\end{document}